\documentclass[runningheads, envcountsame, a4paper]{llncs}
\usepackage[misc]{ifsym}
\usepackage{algorithm}
\usepackage{algorithmic}

\usepackage{graphicx} 
\usepackage{amsmath}
\usepackage{hyperref}
\usepackage{enumitem}
\usepackage{newfloat}
\usepackage{listings}
\usepackage{multirow}

\newcommand{\name}{\texttt{QDiff}}
\begin{document}
\title{Auxiliary Task Guided Interactive Attention Model for Question Difficulty Prediction}
%
%
\author{Venktesh V \Letter \and Md. Shad Akhtar \and
Mukesh Mohania \and
Vikram Goyal}
\tocauthor{Venktesh ~ V, Md. Shad ~ Akhtar,   Mukesh  ~ Mohania,  Vikram ~ Goyal}
\toctitle{Auxiliary Task Guided Interactive Attention Model for Question Difficulty Prediction}
\institute{Indraprastha Institute of Information Technology, Delhi \email{\{venkteshv,shad.akhtar,mukesh,vikram\}@iiitd.ac.in}
}
%
\maketitle              
\begin{abstract}

Online learning platforms conduct exams to evaluate the learners in a monotonous way, where the questions in the database may be classified into Bloom's Taxonomy as varying levels in complexity from basic knowledge to advanced evaluation.  The questions asked in these exams to all learners are very much static. It becomes important to ask new questions with different difficulty levels to each learner to provide a personalized learning experience. In this paper, we propose a multi-task method with an interactive attention mechanism, Qdiff, for jointly predicting Bloom's Taxonomy and difficulty levels of academic questions. We model the interaction between the predicted bloom taxonomy representations and the input representations using an attention mechanism to aid in difficulty prediction. The proposed learning method would help learn representations that capture the relationship between Bloom’s taxonomy and difficulty labels. The proposed multi-task method learns a good input representation by leveraging the relationship between the related tasks and can be used in similar settings where the tasks are related. The results demonstrate that the proposed method performs better than training only on difficulty prediction. However, Bloom's labels may not always be given for some datasets. Hence we soft label another dataset with a model fine-tuned to predict Bloom's labels to demonstrate the applicability of our method to datasets with only difficulty labels. 

\end{abstract}
\keywords{Question difficulty prediction \and Transformers \and Multi-task learning.}
\section{Introduction}
\label{sec:intro}
The academic questions in online learning platforms help the learner evaluate his understanding of concepts. However, serving a static set of questions to all the users is not desirable as all the users do not have the same learning abilities. Hence, there is a need to dynamically adapt to the user's learning profile and accordingly select a question. This would require accurate prediction of the difficulty level of each question so that the system or the academician can choose appropriate questions for the exams.
\begin{table}
\small

\caption{Some samples from our QC-Science dataset.}\label{tab1}
\begin{tabular}{p{8cm}|c|c}
\small
Question text & Difficulty & Bloom's taxonomy \\ \hline \hline
The value of electron gain enthalpy of chlorine is more than that of fluorine. Give reasons. & \multirow{2}{*}{Difficult} & \multirow{2}{*}{Understanding} 
\\ 
\hline
What are artificial sweetening agents? & \multirow{1}{*}{Easy} & \multirow{1}{*}{Remembering} \\\hline
Explain the concept of rotation of Earth. & \multirow{1}{*}{Medium} & \multirow{1}{*}{Understanding} \\ \hline
\end{tabular}

\end{table}
A system for labelling the new questions with appropriate difficulty levels would obviate the need for manual intervention. When the questions are automatically labelled with difficulty levels, it helps in designing adaptive tests where questions in tests are dynamically changed at test time according to the performance in previous questions, tests, or as per the users' capability. For instance, a student who answers the questions in a certain topic like `calculus' correctly is presented with a question of increasing difficulty as the test progresses. This strategy is adopted in online platforms that offer practice for standardized tests like GRE\footnote{https://www.prepscholar.com/gre/blog/how-is-the-gre-scored/}.



Given the advantages of automated difficulty prediction, we propose, \name, a method for predicting the difficulty label of a question that is derived from the difficulty levels denoted as \textit{'easy'}, \textit{'medium'}, or \textit{'difficult'}. We collect a set of academic questions in the Science domain from a leading e-learning platform\footnote{We don't disclose the identity of the source due to the anonymity requirement.}. Bloom's taxonomy provides a mechanism for describing the learning outcomes. The different levels in Bloom's taxonomy as observed in our dataset are \textit{`remembering}', \textit{`understanding'}, \textit{`applying'}, and \textit{`analyzing'} which form the Bloom's labels. The questions are tagged with an appropriate level in Bloom's taxonomy \cite{Gogus2012} and a difficulty level. Some samples from our dataset, named QC-Science, are shown in Table \ref{tab1}. From the collected QC-Science data, we observe that the difficulty level is related to the levels in Bloom's taxonomy as shown in Table \ref{tabbloom}. For instance, in Table \ref{tabbloom} most of the questions tagged with the \textit{`remembering'} level of Bloom's taxonomy are categorized as being \textit{`easy'} questions. 
To verify the strength of association between Bloom's taxonomy and the difficulty levels, we use the \textit{Cramer's V} test since it is best suited for a large sample size. We compute the value V using the formula, 
$V = \sqrt{\dfrac{\chi^2}{n(min(r-1,c-1))}}$
 where $\chi^2$ is the chi-squared statistic, $n$ is the total sample size, $r$ is the number of rows and $c$ is the number of columns.
We obtain a value of \textbf{0.51} for $V$, indicating that there is a strong association between Bloom's taxonomy and the difficulty labels.
Therefore, the Bloom's taxonomy labels can serve as a strong indicator for the difficulty labels and could help in the question difficulty prediction task. 

  \begin{table}
\small
 \centering
 \caption{Distribution of samples across Bloom's taxonomy levels and difficulty levels (contingency table)}\label{tabbloom}
\begin{tabular}{c|c|c|c|c|c}
\multicolumn{1}{c}{}& & \multicolumn{4}{c}{Bloom's Taxonomy} \\ \cline{3-6}
\multicolumn{1}{c}{} & & Analyzing & Applying & Remembering & Understanding\\
\hline \hline
\multirow{3}{*}{\rotatebox{90}{Difficulty}} & Easy &  756 & 1488 &  7146 &  4505\\
& Difficult & 2089 & 2529 & 2518 & 7010\\
& Medium &  585 & 980& 1712 & 2242\\
\hline
\end{tabular}
\vspace{1em}
\end{table}
\paragraph{}
As mentioned in the previous section, we observe a strong association between Bloom's taxonomy labels and difficulty labels. Hence, we propose an interactive attention model to predict the difficulty level and Bloom's taxonomy level jointly for the questions collected for classes VI to XII in the K12\footnote{Kinder-garden to grade-12} education system. 

The Bloom's taxonomy level prediction is considered as an auxiliary task, and the attention weights are computed between the vector representations of the predicted Bloom's taxonomy labels and the input vector representation through the interactive attention mechanism. We use the \textit{hard parameter sharing} approach \cite{ruder2017overview} where the backbone is a Transformer-based \cite{vaswani2017attention} model (like BERT \cite{BERT}) with task specific output layers on top of the backbone network. The conventional multi-task learning methods do not explicitly model the interactions between the task labels and the input. Hence we propose the interactive attention mechanism to explicitly model the interaction between Bloom's taxonomy label and the input, which enables to model the relationship between the tasks better.
We observe that \name\  outperforms the existing baselines as measured by the macro-averaged and weighted-average F1-scores.

\noindent Following are the core technical contributions of our work:
\begin{itemize}[leftmargin=*]
    \item We propose a multi-task learning and interactive attention based approach, \name\ for difficulty prediction, where Bloom's taxonomy predictions are used to determine the input representations using an attention mechanism.
    \item We evaluate the proposed method on the QC-Science dataset. We also evaluate the proposed method on another dataset QA-data \cite{smith2008question} which consists of only difficulty labels. We show that our method trained on QC-Science dataset can be used to soft-label the QA-data dataset with Bloom's taxonomy levels. The experiments demonstrate that our method can be extrapolated to new datasets with only difficulty labels.
\end{itemize}

 Code and data are at \url{https://github.com/VenkteshV/QDIFF_AIED_2022}. 
\section{Related Work}
Question difficulty prediction is an intriguing NLP problem, but it has not been explored to the extent it deserves. A few recent works in difficulty prediction focus on evaluating the readability of the content \cite{yaneva-etal-2017-combining,francois-miltsakaki-2012-nlp}. Authors in \cite{francois-miltsakaki-2012-nlp} proposed to combine classical features like Flesch readability score with non-classical features derived automatically. In  \cite{yaneva-etal-2017-combining}, the authors observed that combining a large generic corpus with a small population specific corpus improves the performance of difficulty prediction.

Another application where the task of question difficulty prediction has been discussed is automated question generation\cite{ha-yaneva-2018-automatic}. In the work \cite{ha-yaneva-2018-automatic}, the authors propose to estimate the difficulty measure as the semantic similarity between the question and the answer options. Similarly, in the work \cite{6644389}, the authors propose a similarity based theory for controlling the difficulty of the questions. 

Prior work has also focused on estimating the difficulty of questions in community question-answering (QA) platforms like StackOverflow \cite{liu-etal-2013-question,wang2014regularized}.
Other studies like \cite{pado-2017-question,nadeem-ostendorf-2017-language}
explore automated grading by estimating the difficulty of academic questions from the Science domain. In the work \cite{nadeem-ostendorf-2017-language}, the authors propose to infer the difficulty of the questions by first mapping them to the concepts.  In the work \cite{xue-etal-2020-predicting}, the authors propose fine-tuning the ELMo \cite{peters-etal-2018-deep} model on 18,000 MCQ type questions from the medical domain for predicting the question difficulty. More recently, in the work \cite{benedetto-etal-2021-application}, the authors propose fine-tuning transformer based language models like BERT for the task of difficulty estimation. 
However, the discussed works do not exploit information from related tasks.

\section{Methodology}
In this section, we describe the proposed approach \name\ for the question difficulty prediction as the primary task and Bloom's taxonomy prediction as the secondary task.  The input to \name\ is a corpus of questions, $C=\{q_1, q_2,..., q_n\}$ where each $q_i$ corresponds to a question along with a difficulty label and Bloom's taxonomy label (skill levels). Since most questions are short texts, we augment the question with the answer as auxiliary information to obtain more semantic information. Hence, we refer to the augmented question as a \textit{`question-answer'} pair in the remainder of the paper. We show that the performance of various methods improves when using the question along with the answer rather than the question text alone. 
\begin{figure*}
\centering
\includegraphics[width=0.75\linewidth]{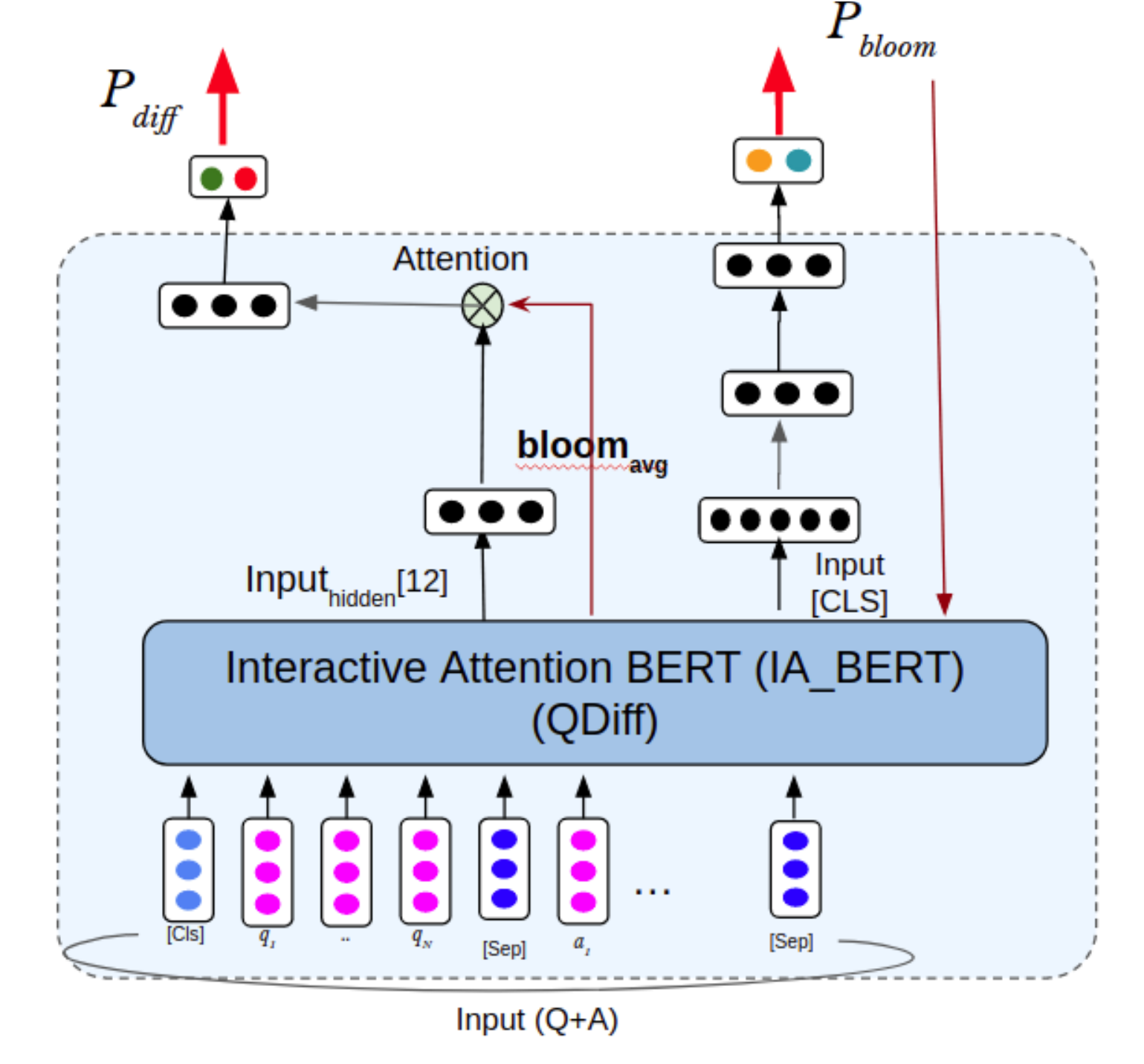}
\caption{\name\ network architecture (middle).}
\label{fig:arch}
\end{figure*}
We obtain contextualized representations for the inputs using BERT \cite{BERT} followed by task specific layers, $H^{\text{\em diff}}$ and $H^{\text{\em bloom}}$ where each tasks specific layer comprises of two linear layers with a non-linearity in between the two layers.

\subsection{Contextualized Input Representations - BERT}
\vspace{-0.14cm}
The academic questions also have \textbf{polysemy} terms that refer to different semantics depending on the context of their occurrence in the input sentence. To tackle the mentioned problem, we use BERT, a transformer-based masked language model, for projecting the input text to the embedding space. 

\textit{Self-attention} is the core of Transformer \cite{vaswani2017attention} model, and BERT utilizes it to obtain better representations.
Self-attention encodes each word in the sentence using \textbf{Q}uery, \textbf{K}ey, and \textbf{V}alue vectors to obtain attention scores, which determines how much attention to pay to each word when generating an embedding for the current word. Mathematically, it is defined as:
\begin{eqnarray}
        Attention(Q,K,V) & = & \dfrac{Softmax(Q * K^T)}{\sqrt{d_k}} *V 
\end{eqnarray}

where, $d_k$ is the dimension of query, key, and value vectors and is used to scale the attention scores, and $K^T$ denotes the transpose of the \textbf{K}ey vector.
\begin{algorithm}[t]
 \small
 \caption{Difficulty prediction}
 \label{algo:main}
 \begin{algorithmic}[1]
 \renewcommand{\algorithmicrequire}{\textbf{Input:}}
 \renewcommand{\algorithmicensure}{\textbf{Output:}}
 \REQUIRE Training set $T \gets $ docs $\{q_1,..q_n\}$, difficulty levels (labels) $y^{\text{\em diff}}$ and bloom's taxonomy labels $y^{\text{\em bloom}}$, test set $S$\\
 \ENSURE  Difficulty levels for the test set $DT$
  \STATE Get input text embeddings , $T_{emb},T_{pooled} \gets BERT(T)$, where $T_{emb}$ represents the set of word embeddings
    \STATE Obtain Bloom's taxonomy level predictions,\\ $ P_{bloom} \gets text\_decode(H^{\text{\em bloom}}(T_{pooled}))$, where $P_{bloom}$ is the text representation of predicted Bloom's label.
    \STATE Obtain the embeddings of the predicted Bloom's taxonomy, \\ $bloom\_emb \gets BERT(P_{bloom})$
    \STATE Obtain average pooled representation of $P_{bloom}$, \\ $bloom\_avg \gets \sum\limits_{i=1}^n\dfrac{ bloom\_emb^{i}}{n}$ where $n$ is the number of subwords in $P_{bloom}$.
    \STATE Compute attention weights,\\ $\alpha_i \gets softmax(f_{attn}(T_{emb}^i, bloom\_avg))$ 
    \STATE Obtain final text representations, $T_{r} \gets \sum\limits_{i=1}^n \alpha_i T_{emb}^i $
  \STATE Obtain difficulty level predictions, $P_{diff} \gets H^{\text{\em diff}}  (T_{r})$ 

  \STATE $\mathcal{L}_{diff} \gets Cross\_entropy(P_{diff},y^{\text{\em diff}})$
    \STATE  $\mathcal{L}_{bloom}\gets Cross\_entropy(P_{bloom},y^{\text{\em bloom}})$
    \STATE $\mathcal{L} \gets \mathcal{L}_{bloom}+\mathcal{L}_{diff}$
    \STATE Fine-tune BERT layers and train the task specific layers to minimize $\mathcal{L}$
 \end{algorithmic} 
 \end{algorithm}
\vspace{-0.07cm}
\subsection{Auxiliary task guided Interactive attention model}
\vspace{-0.07cm}
Based on the strength of association between Bloom's labels and the difficulty labels verified through Cramer's V test (V = \textbf{0.51}), we hypothesize that leveraging Bloom's taxonomy representations to compute input representations using an attention mechanism would lead to better performance in difficulty prediction. The proposed approach would help capture the relationship between the words in the input question and Bloom's taxonomy level, leading to better representations for the task of difficulty prediction as Bloom's taxonomy and difficulty levels are related. Our method also jointly learns to predict both Bloom's taxonomy and the difficulty level, obviating the need for providing Bloom's taxonomy labels at inference time. Figure \ref{fig:arch} shows the architecture of the proposed approach  \name\ . 

    During training, as shown in Algorithm 1, the question-answer pair is first passed through a transformer based language model BERT, to obtain contextualized word representations ($T_{emb}$) and the pooled representation $T_{pooled}$ (step 1). Then the representations are passed to the task specific output layer $H^{\text{\em bloom}}$ to obtain Bloom's taxonomy level predictions (step 2). The vector representations for Bloom's taxonomy prediction are obtained using the \textbf{same BERT} model (step 3). Then the representations of subwords in $P_{bloom}$ are averaged to obtain a fixed 768 dimensional vector representation (step 4). With the input vector representations $T_{emb}$, the attention mechanism generates the attention vector $\alpha_i$ using Bloom's taxonomy representations $bloom\_avg$ (step 5) by
    \begin{eqnarray}
            \alpha_i &= & \dfrac{exp(f_{attn}(T_{emb}^i,bloom\_avg))}{\sum_i^N exp(f_{attn}(T_{emb}^i,bloom\_avg))}  \\
            f_{Attn} & = & \tanh(T_{emb}^i . W_a . bloom\_avg^T + b_a)
    \end{eqnarray} 
    where, tanh is a non-linear activation, $W_a$ and $b_a$ are the weight matrix and bias, respectively.
    
    Then the final input representations are obtained using the attention weights $\alpha_i$ (step 6). The difficulty predictions are then obtained by passing the final input representation $T_{r}$ through the task specific layer $H^{\text{\em diff}}$ (step 7).
    The BERT model acts as the backbone as its parameters are shared between the tasks. The loss function is a combination of the loss for difficulty prediction $\mathcal{L}_{diff}$ and the loss for Bloom's taxonomy level prediction $\mathcal{L}_{bloom}$ (steps 8, 9 and 10). The BERT layers are fine-tuned and the task specific layers are trained to minimize the combined loss (step 11). 

    During the \textit{inference} phase, the contextualized representations are obtained for the test set question-answer pairs as described above. Subsequently, the softmax probability distributions of the difficulty labels are obtained by passing the contextualized representations through the $H^{\text{\em diff}}$ layer. Since the BERT layers are shared between the tasks during training, the contextualized representations obtained at test time improve the performance on the task of difficulty prediction. Since we use an interaction layer with attention where the interaction between the input representations and Bloom's taxonomy label representations are captured, we also call our method as \textbf{IA\_BERT} (Interactive Attention BERT).

\section{Experiments}
In this section, we discuss the experimental setup and the datasets used.
\subsection{Dataset}

\textbf{QC-Science}: We compile the QC-Science dataset from a leading e-learning platform. It contains 45766 question-answer pairs belonging to the science domain tagged with Bloom's taxonomy levels and the difficulty levels. 
 We split the dataset into 39129 samples for training, 2060 samples for validation, and 4577 samples for testing. Some samples are shown in Table \ref{tab1}. The average number of words per question is 37.14, and per answer, it is 32.01. \\
\textbf{QA-data}: We demonstrate the performance of the proposed method on another dataset \cite{smith2008question}. The dataset is labeled only with difficulty labels. We soft-label the dataset with Bloom's taxonomy levels using the Bloom's taxonomy prediction model trained on the \textbf{QC-Science} dataset. We demonstrate that the proposed model, when fine-tuned on a large enough dataset like QC-Science with labels for both the tasks, can be extrapolated to datasets without Bloom's taxonomy labels. The dataset consists of 2768, 308 and 342 train, validation and test samples, respectively. The average number of words per question is 8.71, and per answer, it is 3.96.
\subsection{Baselines}



\noindent We compare with baselines like \textit{LDA + SVM} and \textit{TF-IDF + SVM} \cite{Supraja2017TowardTA}, \textit{ELMo fine-tuning} \cite{xue-etal-2020-predicting}.  We also propose a new baseline.
\begin{itemize}[leftmargin=*]
    \item \textit{TF-IDF + Bloom verb weights (BW)}: In this method the samples are first grouped according to difficulty levels. Then, we extract the Bloom verbs from each sentence using following POS tag patterns: 'VB.*', 'WP', 'WRB'. 
Once the Bloom verbs are extracted, we obtain Bloom verb weights as follows: 

\[Bl\_W = freq(verb, label) * \dfrac{no. \;of \;labels}{n_l}\]

where, $freq(verb, label)$ indicates the number of times a verb appears in a label, and $n_l$ indicates the number of labels that contain the verb. The above operation assigns higher relevance to rare verbs. Then the TF-IDF vector representation of each sentence is multiplied by the weights of the Bloom verbs contained in the sentence. Then for each difficulty level, we obtain the mean of the vector representations of the sentences (centroid). At inference time, the difficulty label whose centroid vector representations is closest to test sentence representation is obtained as output.

\end{itemize}

We also explore the following deep learning approaches.
\begin{itemize}[leftmargin=*]
    \item \textit{Bi-LSTM with attention} and \textit{Bi-GRU with attention and concat pooling \cite{howard2018universal}}.

\item \textit{BERT cascade}: In this method, a BERT (base) model is first fine-tuned on Bloom's taxonomy level prediction followed by the task of difficulty prediction.

\end{itemize}
We also conducted several ablation studies for the proposed architecture \name \hspace{0.1em}.
\begin{itemize}[leftmargin=*]
    \item \textbf{Multi-Task BERT}: In this method, the interaction layer in IA\_BERT is removed, and the model is trained to jointly predict Bloom's taxonomy and difficulty labels.
    \item \textbf{IA\_BERT (B/D) (bloom/difficulty label given)}: In this method, Bloom's taxonomy label is not predicted but rather assumed to be given even at inference time. The model is trained on the objective of difficulty prediction (loss $\mathcal{L}_{diff}$) only. The Bloom's taxonomy label is considered as given when difficulty is predicted and difficulty label is considered as given when the task is Bloom's taxonomy label prediction.
    \item \textbf{IA\_BERT (PB) (pre-trained Bloom model)}: In this method first, a BERT model is fine-tuned for predicting Bloom's taxonomy labels alone, given the input. Then the model's weights are frozen and are used along with another BERT model, which is fine-tuned to predict the difficulty labels.
\end{itemize}
The interaction layer is the same as in IA\_BERT for the last two methods mentioned above.
The HuggingFace library (\url{https://huggingface.co/}) was used to train the models. All the BERT models were fine-tuned for 20 epochs with the ADAM optimizer, with learning rate (lr) of 2e-5 \cite{BERT}. The LSTM and GRU based models are trained using ADAM optimizer and with lr of 0.003.

\begin{table*}[hbt!] 

\caption{Performance comparison for the difficulty prediction and Bloom's taxonomy prediction tasks. $\dagger$ indicates significance at 0.05 level (over Multi-task BERT). D1 - QC-Science, D2 - QA-data }\label{tab:resulst:qa}
\resizebox{1\textwidth}{!}{
\begin{tabular}{l|p{3.5cm}|c|c|c|c|c|c|c|c|c|c|c|c}

 & & \multicolumn{6}{c|}{\bf Difficulty prediction} & \multicolumn{6}{c}{\bf Bloom's level prediction} \\\cline{3-14} & &\multicolumn{3}{c|}{\bf Macro} & \multicolumn{3}{c|}{\bf Weighted} &\multicolumn{3}{c|}{\bf Macro} & \multicolumn{3}{c}{\bf Weighted}\\ \cline{3-14}
\bf & \bf Method & \bf P & \bf R & \bf F1 & \bf P & \bf R & \bf F1 & \bf P & \bf R & \bf F1 & \bf P & \bf R & \bf F1 \\ \hline \hline
& LDA+SVM \cite{Supraja2017TowardTA} & 0.319 & 0.354 & 0.336 & 0.406 & 0.492 & 0.445 & 0.279 &0.266 & 0.267 & 0.320 & 0.360 & 0.338 \\
& TF-IDF + SVM \cite{Supraja2017TowardTA} & 0.471 & 0.415 & 0.440 &0.510 & 0.532 & 0.520 & 0.421 & 0.341 & 0.377 & 0.413 & 0.410 & 0.411 \\
& ELMo \cite{xue-etal-2020-predicting} & 0.466 & 0.403 & 0.432 & 0.495 &0.503 &0.499 & 0.407 & 0.367 & 0.386 & 0.429 & 0.429 & 0.429 \\
& TF-IDF + BW &  0.426 & 0.437 & 0.431 &0.486 &0.429 & 0.456 & 0.330 & 0.346 & 0.338 & 0.364 & 0.344 & 0.342 \\
\cline{2-14}
& Simple rule baseline &  0.359 & 0.402 & 0.379 &0.454 &0.542 & 0.494 & 0.138 & 0.239 & 0.175 & 0.199 & 0.344 & 0.252 \\\cline{2-14}
\textbf{D1}& Bi-LSTM with attention & 0.491 & 0.407 & 0.445 &0.518 &0.529 & 0.523 & 0.487 & 0.419 & 0.450 & 0.497 & 0.481 & 0.489\\
& Bi-GRU with attention & 0.438 & 0.369 & 0.400 & 0.476 & 0.499 & 0.488 & 0.505 & 0.359 & 0.420 & 0.503 & 0.441 & 0.470 \\
& BERT (base) \cite{benedetto-etal-2021-application} & 0.499 & 0.450 & 0.473& 0.530 & 0.550 & 0.539 & 0.484 & 0.459 & 0.471 & 0.494 & 0.502 & 0.498\\
& BERT cascade & 0.494 & 0.454 & 0.473 &0.530 &0.550 &0.539 & 0.470 & 0.441 & 0.455 & 0.486 & 0.486 & 0.486 \\ \cline{2-14}
\cline{2-14}
& Multi-task BERT (ours) &  0.518 & 0.441 & 0.476 &0.538 & 0.556 & 0.547 & 0.490 & 0.439 & 0.463 & 0.497 & 0.499 & 0.498 \\
 & IA\_BERT (\name) (ours) & \bf0.544 & \bf0.447 & \bf0.491$\dagger$ & \bf0.556 & \bf0.564 & \bf0.560$\dagger$ & \bf0.497 & 0.447 & \bf0.471 & \bf0.502 & \bf0.506 & \bf0.504$\dagger$ \\ \hline
& LDA+SVM \cite{Supraja2017TowardTA} & 0.356 & 0.359 & 0.357 & 0.370 & 0.409 & 0.388 & 0.232 &0.205 & 0.218 & 0.580 & 0.655 & 0.615 \\
& TF-IDF + SVM \cite{Supraja2017TowardTA} & 0.518 & 0.487 & 0.502 &0.517 & 0.518 & 0.517 & 0.514 & 0.319 & 0.394 & 0.796 & 0.795 & 0.796 \\
& ELMo \cite{xue-etal-2020-predicting} & 0.635 & 0.623 & 0.629 & 0.654 &0.658 &0.656 & 0.450 & 0.386 & 0.415 & 0.779 & 0.784 & 0.781 \\
& TF-IDF + BW & 0.580 & 0.573 & 0.576 &0.608 &0.581 & 0.594 & 0.312 & 0.338 & 0.324 & 0.705 & 0.646 & 0.674 \\
\cline{2-14}
& Simple rule baseline &  0.236 & 0.341 & 0.279 &0.233 &0.392 & 0.292 & 0.131 & 0.200 & 0.158 & 0.429 & 0.655 & 0.518 \\\cline{2-14}
\textbf{D2}& Bi-LSTM with attention & 0.628 & 0.605 & 0.616 &0.644 &0.655 & 0.650 & 0.430 & 0.357 & 0.390 & 0.765 & 0.766 & 0.766\\
& Bi-GRU with attention & 0.524 & 0.534 & 0.529 & 0.563 & 0.591 & 0.577 & 0.402 & 0.295 & 0.340 & 0.753 & 0.722 & 0.737 \\
& BERT (base) \cite{benedetto-etal-2021-application} & 0.640 & 0.638 & 0.639& 0.661 & 0.660 & 0.661 & 0.455 & 0.404 & 0.428 & 0.814 & 0.822 & 0.818\\
& BERT cascade & 0.662 & 0.666 & 0.664 &0.683 &0.681 &0.682 & 0.437 & 0.401 & 0.418 & 0.816 & 0.827 & 0.821 \\ \cline{2-14}
& Multi-task BERT (ours) &  0.664 & 0.644 & 0.654 &0.681 & 0.687 & 0.684 & 0.389 & 0.365 & 0.377 & 0.799 & 0.804 & 0.802 \\ 

 & IA\_BERT (\name) (ours) & \bf0.684 & \bf0.682 &\bf 0.683$\dagger$ & \bf0.702 & \bf0.708 & \bf0.705$\dagger$ & \bf0.494 & \bf0.420 & \bf0.454$\dagger$ & \bf0.841 & \bf0.830 & \bf0.836$\dagger$ \\

\hline

\end{tabular}}

\end{table*}

\section{Results and Discussion}
The performance comparison of various methods is shown in Table \ref{tab:resulst:qa}. We use macro-average and weighted-average Precision, Recall, and F1-scores as metrics for evaluation. From Table \ref{tab:resulst:qa}, we can observe that most of the deep learning based methods outperform classical ML based methods like TF-IDF + SVM and LDA + SVM. However, we observe that TF-IDF + SVM method outperforms the ELMo baseline\cite{xue-etal-2020-predicting} on the QC-Science dataset. 
It is also evident that the transformer based methods significantly outperform the `TF-IDF + Bloom verbs' baseline. This demonstrates that the contextualized vector representations obtained have more representational power than carefully hand-crafted features for the task of difficulty prediction.
\subsection{Results Analysis}
We provide a detailed analysis of results in this section.
\subsection*{Is the simple rule-based baseline enough?} 
We implement a simple rule-based baseline (Table \ref{tab:resulst:qa}) where the question or answer content is not considered and the difficulty label is predicted based on co-occurrence with the corresponding bloom's label alone. We form a dictionary recording the co-occurrence counts of the bloom's labels and difficulty labels in the training samples as shown in Table \ref{tabbloom}. For each test sample, we look up into the dictionary the entries for the corresponding bloom's taxonomy label of the test sample. Then the difficulty label with maximum co-occurrence count is chosen. We observe that this baseline performs poorly when compared to even other ML baselines from Table \ref{tab:resulst:qa}. This baseline performance demonstrates the need for learning based methods to analyze the given content.

       \begin{table*}[hbt!] 

\caption{Ablation studies for \name \ (IA\_BERT). D1 - QC-Science, D2 - QA-data }\label{tab:resulst:qa:ablation}
\resizebox{1\textwidth}{!}{
\begin{tabular}{l|p{3.5cm}|c|c|c|c|c|c|c|c|c|c|c|c}

 & & \multicolumn{6}{c|}{\bf Difficulty prediction} & \multicolumn{6}{c}{\bf Bloom's level prediction} \\\cline{3-14} & &\multicolumn{3}{c|}{\bf Macro} & \multicolumn{3}{c|}{\bf Weighted} &\multicolumn{3}{c|}{\bf Macro} & \multicolumn{3}{c}{\bf Weighted}\\ \cline{3-14}
 \bf & \bf Method & \bf P & \bf R & \bf F1 & \bf P & \bf R & \bf F1 & \bf P & \bf R & \bf F1 & \bf P & \bf R & \bf F1 \\ \hline \hline
 & IA\_BERT (\name) & 0.544 & 0.447 & 0.491 & 0.556 & 0.564 & 0.560 & 0.497 & 0.447 & 0.471 & 0.502 & 0.506 & 0.504 \\
 \textbf{D1}& IA\_BERT (B/D) & 0.523 & 0.486 & \bf0.503 & 0.554 & \bf0.569 & 0.561 & 0.476 & 0.473  & 0.475 & 0.499 & 0.506 & 0.503 \\
& IA\_BERT (PB) & 0.491 & 0.487 & 0.489 & 0.535 & 0.530 & 0.533 & 0.482 & 0.463 & 0.472 & 0.496 & 0.501 & 0.498 \\ \hline
 & IA\_BERT (\name) & 0.684 & 0.682 & 0.683 & 0.702 & 0.708 & \bf0.705 & 0.494 & 0.420 & 0.454 & 0.841 & 0.830 & 0.836 \\
\textbf{D2}& IA\_BERT (B/D) & 0.682 & 0.688 & 0.685 & 0.702 & 0.696 & 0.699 & 0.458 & 0.429  & 0.443 & 0.837 & 0.842 & 0.839 \\
& IA\_BERT (PB) & 0.642 & 0.641 & 0.641 & 0.667 & 0.652 & 0.659 & 0.449 & 0.424 & 0.436 & 0.825 & 0.825 & 0.825 \\\hline
 \end{tabular}}
\end{table*}

\subsection*{Is the proposed approach better than baselines?}
We observe that \name\ (IA\_BERT), which jointly learns to predict Bloom's taxonomy level and the difficulty level, outperforms all the deep learning (DL) and machine learning (ML) based baselines on both the datasets. It also performs better than BERT (base) \cite{benedetto-etal-2021-application} on the ask of difficulty prediction by advancing weighted F1-score from 0.539 to 0.560 (+3.89\%). It also outperforms Multi-task BERT by \textbf{2.37\%} (weighted F1-score) which can be considered as ablation of the proposed method without the interactive attention mechanism. This demonstrates that in addition to jointly learning to predict labels for both the tasks, the interactive attention mechanism yields better representations leading to improved performance.  We also observe that on the QA-data dataset \name\ (IA\_BERT) outperforms other methods as measured by macro and weighted F1 scores. For Bloom's label prediction, we observe that the IA\_BERT (\name)\ leads to good results (from Table \ref{tab:resulst:qa}) as Bloom's labels are jointly learned and used as signal in attention mechanism. In addition, we also perform an experiment where we use the jointly predicted difficulty labels as signal in the interactive attention mechanism in IA\_BERT for Bloom's label prediction task. We observe that the macro F1-score increases to \textbf{0.479} from 0.471 for QC-Science and also increases the weighted F1-score on QA-data to \textbf{0.852} from 0.836. This demonstrates that both tasks can benefit from each other through the interactive attention mechanism in addition to joint learning. We do not tabulate this result as our focus is difficulty prediction and mention it here for completion.

\subsection*{Are the results statistically significant ?} 
We perform statistical significance test (\textit{t-test}) on obtained outputs. We observe that the results obtained using \name (IA\_BERT) are significant with $p\text{-value} =0.000154$ (weighted-F1) and $p\text{-value} =0.011893$ (macro-F1) for difficulty prediction on QC-Science dataset. We also observe that results are statistically significant on QA-data with $p\text{-value} =0.003813$ (weighted-F1) and $p\text{-value} =0.02248$ (macro-F1) for difficulty prediction.

 \subsection*{Does augmenting question with answer lead to better performance ?}   From Table \ref{tab:resulst:qa}, it is evident that augmenting the question with the answer provides better performance when compared to using the question text alone. When we evaluated on the QC-Science dataset using the question text alone, we observed a drop in performance. For instance, \name\ only yielded a macro-F1 score of 0.455 when using question text alone. The baselines also show a decline in performance, demonstrating that the answer helps provide some context.
\subsection*{What if Bloom's taxonomy labels are randomly labeled for QA-data?}
    We also perform an \textbf{ablation study} on the QA-data dataset by randomly labeling the dataset with Bloom's levels instead of the proposed soft labeling method. We observe that this lowers the performance on the task of difficulty prediction. For instance, \name \ yields a macro F1 score of 0.656 and a weighted F1 score of 0.674 on the difficulty prediction task using the random Bloom's labels. This ablation study supports the significance of the proposed soft labeling method and the interactive attention mechanism as random labels lead to erroneous predictions of difficulty labels.
\subsection*{How does the ablations of IA\_BERT \ \name \ perform?}
We also perform several ablations studies by varying the components of the proposed method. The results are as shown in Table \ref{tab:resulst:qa:ablation}. We observe that using a pre-trained model for Bloom's taxonomy prediction performs poorly as it is not jointly trained on the two related tasks resulting in the error from Bloom's label prediction model propagating to difficulty prediction task through the interactive attention mechanism. Additionally, we observe that directly feeding the Bloom's taxonomy label for the task of difficulty prediction provides gain in performance in certain scenarios as demonstrated by the second ablation in Table \ref{tab:resulst:qa:ablation} for datasets D1 and D2. However, this setting is not possible in real-time scenarios as during inference, the incoming content would not be labeled with Bloom's taxonomy labels. We also observe that the performance of the original IA\_BERT (\name) \ method is very close the mentioned ablation study. This demonstrates that the error propagation from Bloom's label prediction is mitigated in our approach.
\section{Conclusion}
In this paper, we proposed a novel method for predicting the difficulty level of the questions. The proposed method, \name \hspace{0.1em}, leverages an interactive attention mechanism to model the relation between bloom's taxonomy labels and the input text. We observe that \name \hspace{0.2em} and the proposed ensemble construction approach outperforms existing methods. The results also confirm the hypothesis that modeling the interaction between the input and the task labels through an attention mechanism performs better than implicit interactions captured using only multi-task learning. Though question difficulty estimation is subjective, we observe that modeling the interaction between related tasks improves the performance.

\section*{Acknowledgements}

The authors acknowledge the support of Extramarks Education India Pvt. Ltd., SERB, FICCI (PM fellowship), Infosys Centre for AI and TiH Anubhuti (IIITD).
\end{document}